\documentstyle[sprocl]{article}
\def\Journal#1#2#3#4{{#1} {\bf #2}, #3 (#4)}

\def\PRD{{\em Phys. Rev.} D}
\def\APJ{{\em Astrophys. J.}}

\def\AA{{\em Astron. Astrophys.}}

\begin{document}
%
%
\let\ov=\over
\let\lbar=\l
\let\l=\left
\let\r=\right
\def\der#1#2{{\partial{#1}\over\partial{#2}}}
\def\dder#1#2{{\partial^2{#1}\over\partial{#2}^2}}
\def\N{{I\!\!N}}
\def\be{\begin{equation}}
\def\ee{\end{equation}}
%
%

\title{GRAVITATIONAL WAVES FROM ACCRETING NEUTRON STARS \footnote{to appear 
in the Proceedings of the International
Conference on Gravitational Waves: Sources and Detectors, Cascina (Pisa),
Italy --- March 19-23, 1996, Eds.~I.~Ciufolini, F.~Fidecaro  
(World Scientific, in press).}}

\author{S. BONAZZOLA, E. GOURGOULHON}

\address{D\'epartement d'Astrophysique Relativiste et de Cosmologie,\\
UPR 176 C.N.R.S.,\\
Observatoire de Paris, \\
F-92195 Meudon Cedex, France}

\maketitle\abstracts{
  We show that accreting neutron stars in binary systems or in 
Landau-Thorne-Zytkow objects are good candidates 
for continous gravitational wave
emission. Their gravitational radiation is strong enough to be detected
by the next generation of detectors having a typical
noise of $ 10^{-23}{\rm\ Hz}^{-1/2} $. } 

\section{Introduction}

A crude estimate of the gravitational luminosity of an object of mass
$M$, mean radius $R$ and internal velocities of order $V$ can be
derived from the quadrupole formula:
\be	\label{e:quadru,L}
  L \sim {c^5\ov G} \, s^2 \l( {R_{\rm s} \ov R} \r) ^2 
		\l( V \ov c \r) ^6	\ ,
\ee
where $R_{\rm s} := 2 G M / c^2$ is the Schwarzschild radius associated
with the mass $M$ and $s$ is some asymmetry factor: $s=0$ for a
spherically symmetric object and $s\sim 1$ for an object whose shape
is far from that of a sphere. According to formula (\ref{e:quadru,L}),
the astrophysical objects for which 
$s\sim 1$, $R\sim R_{\rm s}$ and $V \sim c$ may radiate a fantastic
power in the form of gravitational waves:
$L\sim {c^5/ G} = 3.6\times 10^{52}$ W, which amounts to
$10^{26}$ times the luminosity of the Sun in the electromagnetic domain!

A neutron star (hereafter NS) 
has a radius quite close to its Schwarzschild radius:
$R \sim 1.5 - 3 \, R_{\rm s}$ and its rotation velocity may reach
$V\sim c/2$ at the equator, so that they are a priori valuable
candidates for strong gravitational emission.
The crucial parameter to be investigated is the asymmetry factor $s$.
It is well known that a uniformly
rotating body, perfectly symmetric with respect to its rotation axis
does not emit any gravitational wave ($s=0$).
Thus in order to radiate gravitationally
a NS must deviate from axisymmetry. Moreover, CW emission 
is possible only if the NS accreets angular momentum from an angular
momentum reservoir. 

Low Mass X-ray Binary systems (LMXB) and High Mass X-rays Binary systems
(HMXB) are a good examples of a NS coupled with an angular momentum reservoir.
These systems
are formed by a NS and an ordinary companion. If the two
stars are close enough, the NS accretes matter (and angular
momentum) from the companion and consequently 
can radiate CW if its axisymmetry is broken.

 The fate of such a system depends on the
mass of the companion. For the HMXB  for
which the companion is a massive star ($\ge 8 M_\odot$ ) and consequentely 
the life time is
quite short ($ \approx 10^6{\rm\ yr} $) 
the companion evolves until when the nuclear fuel 
is exhausted and becomes a supernova. If the binary system is not disrupted
by the explosion, the outcome is a binary pulsar. PSR B1913+16 is a good
illustration of this scenario. If the companion is kicked away by the 
explosion, then the outcome is an isolated pulsar.

 If the mass of the companion is lower than $ 1\, M_\odot $ (LMXB), 
the NS is spun up by the accretion of matter and angular momentum and,
provided the axisymmetry is broken, the NS radiates 
gravitational waves steadily if the accreted
angular momentum is evacuated via gravitational radiation. The light 
companion is evaporated by the electromagnetic emission of the NS and the
final outcome is an isolated millisecond pulsar.

In the intermediate case (mass of the companion between $ 1 M_\odot $
and $ 8 M_\odot $)  the final state is a binary system formed by a white
dwarf and a millisecond pulsar. The important point is that by measuring
the period modulation of the pulsar in a binary system it turns out to be
possible to measure the mass of the NS. The mass of the NS is a
fundamental parameter as will be explained later.

Landau-Thorne-Zytkow objects (LTZO) constitute another example of a NS 
coupled with an angular momentum reservoir. These objects,
introduced by Landau \cite{Landau1938} to explain the stellar source of energy,
have been discussed in details by Thorne and Zytkow \cite{ThoZ77}.
They look as ordinary red supergiant stars, 
the main difference being the core which is a NS 
instead of being white-dwarf like. 
The origin of these objects (if they exist) is
beleived to be a HMXB during the phase in which the NS is orbiting into the
envelop of the companion. Another possible origine are
{\em aborted supernovae}, i.e. supernovae for which the explosion is not 
strong enough to eject the envelop. The NS forming the core of these objects
accrets matter from the envelop at the maximum rate, i.e. at the 
Eddington limit : $ 10^{-8} \, M_\odot {\rm\, yr}^{-1}$. 
If some amount of angular momentum is stored in the envelop, 
the NS is spun up by this accretion. 
The life of a LTZO is about $10^8{\rm\ yr}$, until 
the mass of the NS reaches the maximum value $M_{\rm max}$ 
and the NS collpases into a black hole. 
It must be noticed that the mass of NS, during its life,  
varies between the values of the mass at which the NS is born to the value
of $M_{\rm max}$ that depends on the equation of state (EOS).

\section{Symmetry Breaking Mechanisms} \label{symbreak}

As already said, gravitational waves are radiated by a rotating NS only
if its axisymmetry is broken. Two distinct classes of symmetry breaking 
mechanisms exist: The axisymmetry can be broken spontaneously (via some kind
of instability of the NS) or the axisymmetry can be broken via some external
mechanism. Both cases are pertinent to what follows and therefore will
be discussed in some detail.  

A rotating NS can  break spontaneously
its axial symmetry if the ratio of the rotational kinetic energy
$T$ to the absolute value of the
gravitational potential energy, $|W|$, exceeds some critical value.
When the critical threshold $T/|W|$ is reached, two kinds of instabilities
may drive the star into the non-axisymmetric state:
\begin{enumerate}
\item the {\em Chandrasekhar-Friedman-Schutz instability}
(hereafter {\em CFS instability})  \cite{Chand70}, \cite{FrieS78},
\cite{Fried78} driven
by the gravitational radiation reaction. 
\item the viscosity driven instability \cite{RobeS63}.
\end{enumerate} 

Let us recall some classical results from the theory
of rotating Newtonian homogeneous bodies. 
It is well known that a self-gravitating incompressible fluid rotating rigidly
at some moderate velocity takes the shape of an axisymmetric ellipsoid: 
the so-called {\em Maclaurin spheroid}. At the critical point 
$T/|W| = 0.1375$ in the 
Maclaurin sequence, two families of triaxial ellipsoids branch
off: the {\em Jacobi ellipsoids} and the {\em Dedekind ellipsoids}.
The former are triaxial ellipsoids rotating rigidly about their
smallest axis with respect to an inertial frame, 
whereas the latter have a fixed 
triaxial figure in an inertial frame, with some internal fluid circulation
at constant vorticity   
(see ref.~\cite{Chand69} or \cite{Tasso78} for a review of these classical
results).
The Maclaurin spheroids are dynamically unstable for $T/|W|\geq0.2738$.
Thus the Jacobi/Dedekind bifurcation point $T/|W| = 0.1375$ is dynamically 
stable. However, in presence of some dissipative mechanism such as viscosity
or gravitational radiation (CFS instability) that breaks the circulation
or angular momentum conservation, 
the bifurcation point becomes secularly unstable against 
the $l=2, m=2$ ``bar'' mode. Note also that a non-dissipative mechanism such
as a magnetic field with a component parallel to the rotation axis breaks the
circulation conservation \cite{ChrCK95} and may generate
a spontaneous symmetry breaking. 
If one takes into account only the
viscosity, the growth of the bar mode leads to the deformation 
of the Maclaurin spheroid along a sequence of figures close to some Riemann S 
ellipsoids\footnote{The {\em Riemann S} family is formed by 
homogeneous bodies whose 
fluid motion can be decomposed into a rigid rotation about a principal
axis and a uniform circulation whose vorticity is parallel to the rotation
vector. Maclaurin, Jacobi and Dedekind ellipsoids are all special cases
of Riemann S ellipsoids (for more details, cf. Chap.~7 of ref.~\cite{Chand69}
or Sect.~5 of ref.~\cite{LaiRS93}).}   
and whose final state is a Jacobi ellipsoid \cite{PresT73}.
On the opposite, if the gravitational radiation reaction 
is taken into account but not the viscosity, the Maclaurin spheroid evolves
close to another Riemann S sequence towards a Dedekind ellipsoid
\cite{Mille74}.

 The CFS instability is due to the coupling between the degrees of freedom of
 the star and gravitational waves: the star can loose angular momentum
 (and kinetical energy) via  gravitational radiation. The formation of
waves on the sea when the wind blows is due to an analougous mechanism: in the
frame of reference of the wind, the water looses momentum because of its 
coupling with the atmosphere.
Two conditions must be fulfilled to allow for the growth 
of the CFS instability:
(i) the phase velocity of the pertubation must be less then the rotation
velocity at the equator of the star, (ii) the viscosity must be less than a
threshold value $\mu_{crit} $.

  The first condition is always met: it turns out that the phase velocity of
the gravity waves (the so-called f modes) is $ \propto\ l^{-1/2} $, where $l$
is the ``quantum'' number of the wave in the harmonic functions expansion
(all that in a complete analogy with the sea waves). 
On the contrary, the second condition is hardly fulfilled: the dumping effect
due to the viscosity grows as $ l^2 $. Therefore, taking into the account the
viscosity of nuclear matter, only the mode $l=2$ can survive.
Recent computations  \cite{lindb95} show that
this kind of instability can exist only during a short period in the life
of the star: in fact if the NS is too hot (resp. too cool), 
 the bulk viscosity (resp. the shear viscosity) inhibits the instability.
 Actually the interior of a NS is more complicated: it is superfluid and
type 2 superconductor. Superfluid vortices are coupled with magnetic
fluxoides via their own magnetic field. Vortices and fluxoides are strongly
pinned in the solid crust of the star. All that results in an
effective viscosity higher than the one computed in the absence of magnetic
field. Moreover, any mechanism that tends to rigidify the rotation of star 
(for example the magnetic field) acts against the CFS instability.
From the above it turns out that the CFS instability seems to 
be very unlikely.

 The viscosity driven instability seems to be more promising: in fact, its
rising time decreases when the viscosity increases. 
 The physical mechanism of this instability is very simple: consider the
rotational kinetical energy of the NS at fixed angular momentum $ L $:
$ T = L^2/I $ where $I$ is the moment of inertia with respect to 
the rotation axis. The kinetical energy $T$ decreases if $ I $ increases.
It turns out that for a large enough $ L $, the total energy of the star 
(sum of the kinetical and
gravitational energy) decreases when $ I $ increases. The
natural way to increase $I$ is to let the configuration
to be tri-axial. It is worth to note that the final stellar configuration
is again an equilibrium configuration (in a rotating frame).
The transition between Maclaurin and Jacobi configurations is a real Landau
phase transition of the second order as was showed by Bertin and Radicati 
\cite{BertR1976}. The reader can find more details in our lecture on the 
subject at Les Houches School \cite{BonaG96}. 

The main problem is that this instability can work, as already said, only
if the NS rotates fast enough. The maximun angular velocity of a rigidly
rotating star is achieved when the velocity at the equator is equal to the
Keplerian velocity. It turns out that if the EOS of the fluid forming the 
star is too soft, the Keplerian velocity is less 
than the critical velocity
for which the axisymmetry breaks. For a polytropic EOS 
($P \propto \rho^{\gamma} $) and in the Newtonian theory, 
$\gamma$ must be greater than $ \gamma_{\rm crit} = 2.238 $
(ref.~\cite{James64}, \cite{BonFG96}, \cite{SkinL96}).

\section{Results for realistic equations of state}

Recently, we have generalized the above results to the existing
``realistic'' EOS in a General Relativistic frame
\cite{BonFG96}. Table~\ref{t:resu,EOS} shows the results:
among the 12 EOS taken under consideration, five are stiff enough to 
allow for the transition toward a 3-D configuration.
In table~\ref{t:resu,EOS}, the EOS are labeled by the following 
abbreviations: PandN refers to the
pure neutron EOS of Pandharipande \cite{Pandh71}, 
BJI to model IH of Bethe \& Johnson \cite{BethJ74}, 
FP to the EOS of Friedman \& Pandharipande \cite{FrieP81}, 
HKP to the  $n_0 = 0.17\ {\rm fm}^{-3}$ model of Haensel et al. \cite{HaeKP81},
DiazII to model II of Diaz Alonso \cite{Diaz85}, 
Glend1, Glend2 and Glend3 to respectively the case 1, 2, and 3 of 
Glendenning EOS \cite{Glend85},
WFF1, WFF2 and WFF3 to respectively the 
${\rm AV}_{14}+{\rm UVII}$, ${\rm UV}_{14}+{\rm UVII}$ and 
${\rm UV}_{14}+{\rm TNI}$ models of Wiringa et al. \cite{WirFF88}, 
and WGW to the $\Lambda_{\rm Bonn}^{00}+{\rm HV}$ model of Weber et al.
\cite{WebGW91}.   

\begin{table}
\centerline{
\begin{tabular}{lllllll}
\hline\noalign{\smallskip}
  $\displaystyle{{\rm EOS}\atop \ }$  	&
  $\displaystyle{{M_{\rm max}^{\rm stat}\atop [M_\odot]}}$ 	&
  $\displaystyle{{M_{\rm max}^{\rm rot}\atop [M_\odot]}}$ 	&
  $\displaystyle{{P_{\rm K}\atop [{\rm ms}]}}$ 	&
  $\displaystyle{{P_{\rm break}\atop [{\rm ms}]}}$ 	&
  $\displaystyle{{H_{\rm c,break}\atop \ }}$ 	&
  $\displaystyle{{M_{\rm break}\atop [M_\odot]}}$ 	\\
 \noalign{\smallskip}
\hline\noalign{\smallskip}
  HKP	  & 2.827 & 3.432 & 0.737 & 1.193 & 0.168 & 1.886 \\
  WFF2	  & 2.187 & 2.586 & 0.505 & 0.764 & 0.292 & 1.925 \\
  WFF1 	  & 2.123 & 2.528 & 0.476 & 0.728 & 0.270 & 1.742 \\
  WGW  	  & 1.967 & 2.358 & 0.676 & 1.042 & 0.170 & 1.645 \\
  Glend3  & 1.964 & 2.308 & 0.710 & \multicolumn{3}{c}{stable} \\
  FP	  & 1.960 & 2.314 & 0.508 & 0.630 & 0.412 & 2.028 \\
  DiazII  & 1.928 & 2.256 & 0.673 & \multicolumn{3}{c}{stable} \\
  BJI	  & 1.850 & 2.146 & 0.589 & \multicolumn{3}{c}{stable} \\
  WFF3    & 1.836 & 2.172 & 0.550 & 0.712 & 0.327 & 1.919 \\
  Glend1  & 1.803 & 2.125 & 0.726 & \multicolumn{3}{c}{stable} \\
  Glend2  & 1.777 & 2.087 & 0.758 & \multicolumn{3}{c}{stable} \\
  PandN   & 1.657 & 1.928 & 0.489 & \multicolumn{3}{c}{stable}\\
\noalign{\smallskip}
\hline\noalign{\smallskip}
\end{tabular}
}
\caption[]{\label{t:resu,EOS}
Neutron star properties according to various EOS: $M_{\rm max}^{\rm stat}$
is the maximum mass for static  configurations, 
$M_{\rm max}^{\rm rot}$ is the maximum mass for rotating stationary 
configurations, $P_{\rm K}$ is the corresponding Keplerian period,
$P_{\rm break}$ is the rotation period below which the symmetry breaking occurs, 
$H_{\rm c, break}$ is the central log-enthalpy at the bifurcation point 
and $M_{\rm break}$ is the corresponding gravitational mass.  
The EOS are ordered by decreasing values of $M_{\rm max}^{\rm stat}$.
}
\end{table}

From the above results 
it appears that only NSs
whose mass is larger than
$1.64\, M_\odot $ meet the conditions of spontaneous symmetry breaking via
the viscosity-driven instability. The above minimum mass is quite below
the maximum mass
of a fast rotating NS for a stiff EOS ($3.2\ M_\odot $ \cite{SaBGH94}). 
Note that the critical period at which
the instability happens ($P=1.04$ ms) is not far from the lowest observed one
($1.56$ ms).  The question that naturally arises
is: do these heavy NSs exist in nature ? 
Only observations can give the
answer; in fact, the numerical modelling of a supernova
core and its collapse \cite{Mulle96} cannot yet provide us with a reliable 
answer.
 The maximum critical rotation period ($ 1.2 {\rm\ ms}$) 
at which the the instability
appears is compatible with the rotation period of the fastest known pulsar
($1.56 {\rm\ ms}$);
moreover the age of these pulsar spans between $10^7$ and $10^9 {\rm\ yr}$. 

  The real problem is the minimum mass, $ 1.64 M_{\odot} $, for the
triaxial instability to develop.  
This is not in very good agreement with the measured masses
(all in binary systems) \cite{ThAKT93}, except for PSR~J1012+5307
which appears to be a heavy NS: $1.5 \, M_\odot < M < 3.2 \, M_\odot$
\cite{KerBK96}.  
Four NS masses (all in binary radio pulsars) are known 
with a precision better than $10\% $  and they turn out to be around
$1.4\ M_\odot $ \cite{ThAKT93}. 
Among the X-ray binary NSs, two of them seem to have a higher mass:
4U 1700-37 and Vela X-1 
($1.8\pm 0.5\ M_\odot $ and $1.8\pm 0.3\ M_\odot $ respectively). 
These objects show that NSs in binary systems may
have a mass larger than  $1.64\ M_\odot  $.

A natural question that may arise is: why do X-ray binary NSs,
which are believed to be the progenitors
of binary radio pulsars, have a mass larger than the latter ones ?
We have not yet any reliable answer to this question. 
A first (pessimistic) answer is
that the measurements of X-ray NS masses are bad 
(compare the error bars
of the masses of the binary radio pulsars with the ones of the X-ray binaries 
in Fig.~3 of ref.~\cite{ThAKT93}),
and consequently not reliable. Actually it should be noticed that 
the error bars of the X-ray pulsars do not have the
same statistical meaning as the error bars of the binary radio pulsars
\cite{Lindb96}:
they give only the extremum limits of NS masses
in the X-ray binary. Consequently $1.4\ M_\odot $ is not incompatible with
these masses. 

A related question arises naturally: why are the observed masses 
of millisecond radio pulsars almost identical ? Following the standard model, 
a millisecond radio pulsar is a recycled NS, spun up by 
the accretion of mass and angular momentum from a companion. The observed
mass and angular velocity are those of the end of the accretion process.
Consequently the  accreted mass depends on the history of the system and 
on the nature
of the companion. By supposing ``per absurdo'' that all NSs are born
with the same mass, it is difficult to understand why the accreted mass is 
the {\em same} for all NSs. 
A possible answer is that this could result from some 
observational selection effect. For example, suppose that accreted
matter quenches the magnetic field, it is then easy to imagine that the final 
external magnetic field depends on the mass of the accreted plasma.
If the accreted mass is large enough, the magnetic field can be lower 
than the critical value for which the pulsar mechanism works.  
On the contrary, if the accreted mass is quite small,  
the magnetic field is large  and the life time of the 
radio pulsar phase is shorter and consequently more difficult to observe.

\section{Detectability}

If the nuclear matter EOS is stiff enough and 
accreting NS in binary systems have a mass large enough for 
the symmetry breaking to take place, accreting NS are efficient
gravitational wave emitters. It is very easy to compute the amplitude of
the emitted gravitational waves. By equating the rate of the accreted
angular momentum to the rate of radiated angular momentum 
one obtains \cite{Wagon84}
\be \label{e:h(Fx)}
h=1.3\times 10^{-27} \left(\frac{1 {\rm\ kHz}}{\nu} \right)
\left( \frac{F_X}{10^{-8} {\rm\ erg\,  cm}^{-2}{\rm\,  s}^{-1}} \right)^{1/2}
\ee
where $ h $ is the strain of the gravitational wave, $\nu$ the rotation 
frequency of the source, $ F_X $ the X-ray flux received on Earth.
  Note that the distance of the source does not appear in the above formula. 
 From (\ref{e:h(Fx)}), the signal-to-noise ratio $S/N$ can be easily computed
in terms of the observation time $ T $ and the sensitivity of the
detector $B$. For the brighest X-ray source, Sco X-1 
($ F_X = 2\times 10^{-7} {\rm\ erg\, cm}^{-2}{\rm\,  s}^{-1}$), we obtain 
\be
 \left( {S\over N} \right) _{\rm Sco\ X-1} = 
	\left(\frac{0.17}{B/(10^{-23}{\rm\ Hz}^{-1/2})} \right)
	\left(\frac{1{\rm\ kHz}}{\nu}
	\right)\left(\frac{T}{1 {\rm\ day}} \right)^{1/2}
\ee
From the above formula, we see that one month of observation is 
sufficient to obtain $ S/N =1 $ with a detector of the 
$10^{-23} {\rm\ Hz}^{-1/2}$ class.
 This is however a misleading result: in fact, because the frequency of the
CW emission is not known, a signal-to-noise of about 7 is required 
in order to have a detection with a confidence level equivalent to the
ordinary $ 3 \sigma $ criterium. This means that 2.5 years of observation time
with 
{\em one} detector are necessary to detect the gravitational radiation
emitted by rotating NS. With 3 detectors (e.g. 2 LIGO + VIRGO) the situation 
appears more favorable: 10 months of observation time instead of 30. Moreover
a less naive strategy can be used to couple the 3 detectors; we do not discuss 
this possibility here.

LTZO objects are also good candidates. The radiation mechanism is analogous to
that of the accreting binary sources: the NS forming the core is spun
up by the accreted  angular momentum from the envelop. 
The main avantage of these sources is that the mass range of the 
inner NS spans from the initial mass of the NS ($\sim 1.4\ M_\odot$) up to the 
critical mass ($ \geq 2\ M_\odot $). The drawback is that we do not if these 
objects  exist.    

Finally, note that a deformation (ellipticity) 
as small as $ \approx 10^{-8} $ is sufficent
to radiate the accreted angular momentum at the Edington mass accretion rate
($ 10^{-8} \, M_\odot{\rm\, yr}^{-1}$).
A question naturally arises: do there exist any other mechanism
able to deform the NS by a such a amount ? No alined magnetic field can do
the job. The accreted matter is funelled by the magnetic 
field onto the crust of the NS, and spreads out on the surface, 
but magnetic field acts  as a magnetic brake for this process.  
The efficiency of this magnetic braking depends
on the conductivity of the plasma and on the strength of the magnetic field.
A rough estimation of the typical spreading  time $ \tau $ of the accreted
plasma on the surface of the NS gives $ \tau \gg 1{\rm\ yr}$. This means that
3-D asymetries can be larger than $ 10^{-8} $ for the accreting rate of
$ 10^{-9} - 10^{-8}\ M_\odot {\rm\, yr}^{-1}$. 
The above encouraging result is correct
only if the effecive conductivity of the plasma is equal to the microscopic
one. Indeed plasma instabilities can reduce the effective conductivity
by orders of magnitude. The most dangerous of them is the instability generating
the reconnection of the magnetic field lines. Fortunately, no $ X $ or $ O $ 
point exists in the magnetic field configuration. Therefore this kind of
instability seems to be excluded. More investigation is needed to clarify this
important question (the Authors thank Prof. E.~Spiegel and Dr. A.~Mangeney
for illuminating discussions on this point).

\section{CONCLUSION}

  Accreting NS in binary systems or in LTZOs can be good gravitational
CW emitters. Their positions on the sky are known, therefore  data can be
easily reduced to the solar system barycentric frame and the Doppler shift 
induced by the motion of the Earth can be properly taken into account. 
The amplitude of the predicted gravitational waves is large enough 
to be detected with
the $ 10^{-23}{\rm\ Hz}^{-1/2} $ class of detectors. Positive detection will give
to us important informations on the equation of state of the nuclear matter
in NS. The proof of existence of the LTZOs will be a major discovery
leading to  important informations on the stellar evolution during the
common envelop phase.

\end{document}